\documentclass[prc,onecolumn,tightenlines,12pt]{revtex4}
\usepackage{graphicx}
\usepackage{dcolumn}
\usepackage{amsmath}
\begin{document}

\title{Decomposition of spectral density in 
individual eigenvalue contributions}

\author{O. Bohigas$^{1}$ and M. P. Pato$^{1,2}$}

\affiliation{$^{1}$CNRS, Universit\'e Paris-Sud, UMR8626, \\
LPTMS, Orsay Cedex, F-91405, France \\
$^{2}$Instituto de F\'{\i}sica, Universidade de S\~{a}o Paulo\\
Caixa Postal 66318, 05314-970 S\~{a}o Paulo, S.P., Brazil}

\begin{abstract}

The eigenvalue densities of two random matrix ensembles, 
the Wigner Gaussian matrices and the Wishart covariant matrices, 
are decomposed in the contributions of each individual eigenvalue
distribution. It is shown that the fluctuations of all eigenvalues,
for medium matrix sizes, are described with a good precision 
by nearly normal distributions.

\end{abstract}

\maketitle

\section{Introduction}

Borrowing from J. Wishart the random matrices used by
statisticians 
to construct their ensemble of covariant matrices\cite{Wishart},
E. Wigner, in the late fifties, introduced
the ensemble of Gaussian matrices of random matrix theory (RMT)
\cite{Porter}. 
Spectral properties of these two ensembles are 
characterized by correlations generated by the 
repulsion among the levels and their properties are
directly connected to two classical polynomials, the Hermite ones, 
in the Wigner case\cite{Mehta}, and the Laguerre ones, 
in the Wishart case. Both 
ensembles, and specially Wigner's, have proved to be 
important matrix models. By the same time, 
M. Girardeau proved the mapping theorem that states that
in 1D the properties of a gas of impenetrable bosons  
are the same of a gas of fictitious spinless fermions
\cite{Gira}. In the limit of negligible size, the only 
interaction among the atoms that remains is that it is forbidden for 
two atoms to occupy the same site. This generates a repulsion 
among them in an entirely similar way of what occurs among the 
eigenvalues. Placed in a harmonic trap, it follows 
from this equivalence that the ground
state wave function of the atomic system is just the 
joint probability distribution of RMT 
eigenvalues\cite{Kolo,Gira2}. Therefore, under these 
conditions, the atoms of the unidimensional boson gas constitute 
a physical realization of eigenvalues of the matrix ensemble.   
 
In the last decades, there has been in RMT studies a great
interest in the behavior of individual eigenvalues at the edge of 
spectra. This interest followed the many applications that 
the probability distribution 
derived by C. Tracy and H. Widom which describes
the behavior of eigenvalues at the border of the Gaussian ensemble
\cite{TW} have found. The distributions of the largest and the smallest
eigenvalues of Wishart matrices have also been 
derived\cite{Johnstone}. A salient feature of these edge distributions
is their asymmetry that can be understood as an effect of the unbalanced 
repulsion that eigenvalues at the border are submitted from the other
eigenvalues. Here we are interested in investigating the behavior 
of eigenvalues not only at the edge but also at the bulk of the spectrum.  
This question has been aroused some years ago in connection with the
behavior of eigenvalues of the two-body random ensemble as
compared with the RMT ones\cite{Flores}. 
In the boson gas terms, this decomposition is equivalent to determine 
how individual atoms behave.

First we remark that, in general, the density of a set of random
variables can be expressed as a sum of the individual distributions 
of each variable considered in an ordered sequence. Of course, 
this just translates the fact that a variable found at a given position 
has to be or the first, or the second, or the third, and so on, 
of the ordered sequence. Therefore, if the distribution of
each variable is determined, a decomposition of the
density follows. Intuitively, it is reasonable to expect that, 
at the bulk of the 
spectrum, if not exactly at least approximately in the case of
large matrices, eigenvalue fluctuations will be normally
distributed and, in fact this has been proved in the limit when 
the size of the matrices goes to infinity\cite{Jonas,Sean}. 
As a consequence, the exact expression of the eigenvalue density 
should allow a decomposition in terms of nearly Gaussian
distributions. It is our purpose to show that this decomposition 
exists and to determine its parameters. This will complement the 
exact results obtained in the asymptotic limit when matrix sizes
go to infinity\cite{Tao}. To do this, we note that
the exact density of the eigenvalues of these
two ensembles show tiny fluctuations around the average density. 
These wiggles correspond to the peaks of the individual eigenvalue 
distributions, that is to the average positions of the eigenvalues.
To find these positions we separate in the eigenvalue density, 
a smooth leading order term of the fluctuating term which vanishes 
in the limit of large matrix size. The locations 
of the wiggles are then determined in the fluctuating term. 
Comparing then this term with the one of 
the same order in the asymptotic of the Gaussian decomposition, 
the dependence of the Gaussian variances on their positions along 
the spectrum is determined.  

The Wigner and the Wishart ensembles are characterized by Dyson 
index $\beta$ that takes the values $1,2,4$ for the three symmetry 
classes,respectively,  the orthogonal (GOE) of real symmetric matrices, 
the unitary (GUE) of complex hermitian matrices and the 
symplectic (GSE) whose elements are real quaternion numbers. 
A $\beta$-ensemble has been proposed that generalizes, for arbitrary 
positive values of $\beta,$ the Wigner and the Wishart 
ensembles\cite{Edelman}. In particular, in the limit 
$\beta\rightarrow\infty,$ the spectrum gets frozen with eigenvalues 
located at the position of the zeros of the Hermite 
(Laguerre, in Wishart case) polynomials. 
As $\beta$ decreases, the eigenvalues start
to vibrate around those positions. Using perturbation theory, 
a Gaussian decomposition of the 
eigenvalue density for large but finite $\beta$ was derived 
for both ensembles\cite{Edelman2}. 
Here we are extending this decomposition for
small values of $\beta.$ For large $\beta$ individual distributions 
do not overlap while in our case there is overlapping 
between neighboring distributions. In the opposite limit 
$\beta\rightarrow 0,$ the eigenvalues become a set of 
uncorrelated variables and, in this case, there is strong overlapping
among the individual distributions. 

\section{Uncorrelated variables}

Starting with the case of uncorrelated variables, we consider a set 
of $N$ independent and identically distributed (i.i.d.) 
variables $x_i$ with $i=1,2,...,N$ 
uniformly distributed in the interval $(-N/2,N/2).$  By
simple probabilistic argument, 
the density probability of finding a variable at $t$ with $n$ others greater and
the $N-1-n$ others smaller than it, is 

\begin{equation}
F(n,t)=\frac{(N-1)!}{n!(N-n-1)!}\left(\frac{1}{2}-\frac{t}{N}\right)^n  
\left(\frac{1}{2}+\frac{t}{N}\right)^{N-n-1} .
\label{3}
\end{equation}  
With $n=0,1,,...,N-1,$ this family of beta distributions  gives an exact 
description of the order statistics of the set of 
variables in which the 
$n$th function, $F(n,t),$ corresponds to the density 
distribution of the $(n+1)$th largest variable.
Immediately, we verify that summing up
all of them, the unit density is recovered as it should, namely  
$\rho(t)=\sum F(n,t)=1.$
From (\ref{3}), it follows that these distributions have first moment  
and variance given by 
\begin{equation}
\bar t =\frac{N(N-1-2n)}{2(N+1)} \label{21a}
\end{equation}
and 

\begin{equation}
\sigma^2=\frac{(n+1)(N-n)N^2}{(N+2)(N+1)^2} , \label{21}
\end{equation}
respectively.

By taking the limit $N\rightarrow \infty$ keeping $n$ fixed, the order 
statistics of the largest variables is obtained. In this case
it is more convenient to express the distributions using as variable 
the distance $y=t-N/2,$ to the right border in terms of which, 
the distributions $F(n,y)$ converge, for large $N,$ to

\begin{equation}
F(n,t)=\frac{(-y)^n}{n!}\exp(y).  \label{13}
\end{equation}  
This set of functions are known to give the density distributions of the 
largest variables of an i.i.d. sequence with a compact
support\cite{Coles}. 
In particular, the first one, $F(0,y)=\exp(y),$ is  
the Weibull distribution\cite{Coles,Weibull}. From (\ref{13}), 
we find that on the average the $(n+1)$th variable is located at the position
${\bar y}=-(n+1).$

Consider now the situation in which, in the same limit of a large number
of variables, the ratio $\frac{n}{N}$ is kept fixed.
In this case, the beta distributions converge to the Gaussian

\begin{equation}
F(n,t)=\frac{1}{\sigma\sqrt{2\pi}}\exp\left[-\frac{(t-\bar t )^2}
{2\sigma^2}\right]. \label{12}
\end{equation} 
that give the order statistics of the variables at the bulk of the
sequence. Taking the limit
of large $N$ in the expressions (\ref{21a}) and (\ref{21}) the
first moment and the variance become

\begin{equation}
\bar t =\frac{(N-1)}{2}-n
\end{equation}
and 

\begin{equation}
\sigma^2=\frac{n(N-n)}{N} . \label{22}
\end{equation}
The above equation shows that the variance of the individual 
distributions at the bulk scales as $\sigma\sim \sqrt{N}$.
The case of i.i.d. variables with an arbitrary symmetric 
density distribution $\rho(x),$ can be mapped into the above 
case by the transformation 
$t=\int_{0}^{x}dx^{\prime}\rho(x^{\prime}).$ The  density distribution 
of the $(n+1)$th variable of the ordered sequence is then

\begin{equation}
F(n,x)=\rho(x)\frac{(N-1)!}{n!(N-n-1)!}\left[\frac{1}{2}-\frac{t(x)}{N}
\right]^n  
\left[\frac{1}{2}+\frac{t(x)}{N}\right]^{N-n-1} .
\end{equation}  
In Fig. 1, it is shown the individual contribution decomposition
for a set of $N=20$ i.i.d. variables sorted from the common Gaussian 
distribution

\begin{equation}
\rho(x)=\frac{N}{\sqrt{\pi}} \exp(-x^2 ).
\end{equation}  
In this case

\begin{equation}
t(x)= \frac{N}{2} \mbox {erf} (x),
\end{equation}  
where $\mbox {erf} (x)$ is the error function.
  
It is seen that there is a strong overlapping among the individual
distributions that reflects the dependence of the variances with the 
square root of the number $N$ of variables. The largest variable 
is expected to be distributed according to the Gumbel density 
distribution\cite{Gumbel}

\begin{equation}
F_{Gumbel} (z)= \exp [-\exp (-z) ] \exp (-z)  \label{61}
\end{equation}
in the new variable 

\begin{equation}
z= -\ln[ \frac{N}{2}  \mbox {erfc} (x )] ,
\end{equation}
where  $\mbox {erfc} (x)$  is the complementary error function. Indeed,
one can see that this is true from the very good agreement exhibited 
in Fig. 2, in which the largest variable in a sequence of $N=100$ 
Gaussian variables is
compared against the above Gumbel density distribution, Eq. (\ref{61}). 

\section{Correlated eigenvalues}

The Wigner and the Wishart ensembles belong to a class of ensembles
whose joint probability density distribution of the eigenvalues has the form

\begin{equation}
P(x_1,x_2,...,x_N)=K_N \exp\left[-\frac{\beta}{2}\sum_{k=1}^N 
V(x_k)\right]
\prod_{j>i}\mid x_j-x_i\mid^\beta \label{1001} ,
\end{equation}
where $N$ is the number of eigenvalues, $\beta$ is Dyson index and
$K_N$ is a normalization constant. In (\ref{1001}), $V(x)$ is a confining
potential which makes the above distribution normalizable by
keeping the eigenvalues inside a potential. The exact expression for the 
eigenvalue density obtained integrating all variables but one is
given by\cite{Mehta}

\begin{equation}
\rho(x)=\exp[-V(x)]\sum_0^{N-1} P_n ^2(x),\label{1002}
\end{equation}
where the $P_n (x)$ are normalized polynomials orthogonal with respect to the
weight $\exp[-V(x)]$. In fact, this is the density of the unitary
ensemble, $\beta =2 $, but extra terms have to be added in the case of the
orthogonal and the symplectic ensembles. 
The above sum can be reduced to the contribution of
the last term using the Christoffel-Darboux relation 

\begin{equation}
\sum_0^{N-1} P_n (x) P_n (y)=\frac{k_{N-1}}{k_{N}}
\frac{P_N (x)P_{N-1} (y)-
P_N (y)P_{N-1} (x)}{x-y},   \label{1002f}
\end{equation}
where $k_n$ is the highest coefficient of $P_n(x),$ in the limit 
$y\rightarrow x,$  (\ref{1002f}) becomes

\begin{equation}
\sum_0^{N-1} P_n ^2 (x)=\frac{k_{N-1}}{k_{N}}
\left[ P_N ^{\prime} (x)P_{N-1} (x)-
P_N (x) P_{N-1} ^{\prime} (x) \right]. \label{202q}
\end{equation}
The density described by Eq. (\ref{1002}), in the case 
we are interested in of Wigner's and Wishart's ensembles, has a central
part with wiggles separated by inflection points, namely where the
curvature changes sign, 
from the decaying tails at the edges.
In the following, we use Eq. (\ref{202q}) to show that, in the central
bulk of the spectra, the density for these two ensembles can be written as
a sum

\begin{equation}
\rho(x)=\rho_s (x) +  \rho_f (x)
\end{equation}
of a smooth leading term, $\rho_s (x),$ and a fluctuating term, $\rho_f (x),$
which vanishes in the limit of large matrices. From this decomposition
of the density in smooth and fluctuating terms, the parameters which
define individual eigenvalue distributions are extracted. However,
this procedure can not be used beyond the inflection point that
is for the first and the last eigenvalues distributions. But, we
remark that for these extreme eigenvalues, the tail of their densities
coincides with the density itself and does not need to be calculated.
With this procedure, we are able to obtain the individual distribution
of all eigenvalues.

\subsection{Wigner ensembles}

The joint distribution of the eigenvalues of the Gaussian random
matrices for the orthogonal, unitary and symplectic ensembles is

\begin{equation}
P(x_1,x_2,...,x_N)=Z_N^{-1} \exp\left(-\frac{\beta}{2}\sum_{k=1}^N x_k^2\right)
\prod_{j>i}\mid x_j-x_i\mid^\beta ,
\end{equation}
where

\begin{equation}
Z_N(\beta)=(2\pi)^{N/2}\beta^{-N/2 \, -\beta N(N-1)/4}
\prod^{N}_{j=1}\frac{\Gamma
\left(
1+j\beta/2
\right)}
{\Gamma
\left(
1+\beta/2
\right)} .
\end{equation} 
Therefore, from Eq. (\ref{1001}), the confining potential for this 
ensemble is the parabola $V(x)=x^2/2$ and the orthogonal polynomials 
are the Hermite polynomials $H_n (x)$. 

\subsubsection{Unitary ensemble ($\beta =2$)}

The eigenvalue density is\cite{Mehta} 

\begin{equation}
\rho(x)=\sum_0^{N-1} \phi_n ^2(x) ,\label{101}
\end{equation}
where 

\begin{equation}
\phi_n (x)= \frac{\exp(-x^2/2)H_n (x)}{\sqrt{n!2^n \sqrt{\pi}}}
\label{101b}
\end{equation}
satisfies the equation 

\begin{equation}
\frac{d^2 \phi_n (x)}{dx^2}+\left(2n+1-x^2\right) \phi_n (x)=0 .
\end{equation}
which, in appropriate units, is the Schr\"{o}dinger equation
for the quantum harmonic oscillator. 

To be able to later calculate individual distributions of i.i.d. random
variables with the above density, Eq. (\ref{101}), we define the 
counting function

\begin{equation} 
N(x)=\int^x_0 d x^{\prime}\rho(x^{\prime})  . 
\end{equation}
To calculate this function the recurrence relation

\begin{equation}
\phi_n (x)=\sqrt{\frac{2}{n}}x\phi_{n-1}(x)-\sqrt{\frac{n-1}{n}}
\phi_{n-2}(x)     \label{101z}       
\end{equation}
which follows from the known  recurrence relation

\begin{equation}
H_n (x)=2xH_{n-1}(x)-2(n-1)H_{n-2}(x)
\end{equation}
of the Hermite polynomials can be used. Integrating the square of  (\ref{101z}) and
using the relation
  
\begin{equation}
\phi_n^{\prime} (x)=-x\phi_{n}(x)+\sqrt{2n}\phi_{n-1}(x) ,
\end{equation}
derived taking the derivative of  (\ref{101b}) together with

\begin{equation} 
H_n^{\prime}= nH_{n-1} , \label{101v} 
\end{equation}
we deduce the recursion relation 

\begin{equation}
\int_0^x dt \phi_n^2 (t) = -\frac{x}{n}\phi_{n-1}^2 (x)+\frac{1}{n}
\int_0^x dt \phi_{n-1}^2 (t) +\frac{n-1}{n} \int_0^x dt \phi_{n-2}^2 (t)
\end{equation}
which provides an efficient way to calculate the counting function
starting from the ground state function $\phi_0 (x)$.

Before analyzing the case of large matrices, it is instructive to consider 
the simple case of matrices of size $N=2$ for which Eq. (\ref{101}) gives   

\begin{equation}
\rho(x)=\frac{\exp(-x^2)}{\sqrt{\pi}} (1+2x^2)\label{1012} .
\end{equation}
The probability of the largest eigenvalue be less than a value $x$ is obtained by 
integrating the joint distribution $P(x_1 , x_2 )$ in the two
variables $x_1$ and $x_2$ from 
$-\infty$ to $x$. Taking then the derivative of this probability,
we find that the density distribution of the largest eigenvalue is 

\begin{equation}
F(0,x)=\frac{\exp(-x^2)}{2\sqrt{\pi}}\left[(1+2x^2)(1-\mbox{erf}(x))-
\frac{2x\exp(-x^2)}{\sqrt{\pi}}\right] .
\end{equation}
For the smallest eigenvalue distribution it is simpler to determine it
by taking the difference $\rho(x)-F(0,x).$ For the uncorrelated
case (see previous section), with $n=0,1$ and $\rho(x)$ 
given by(\ref{1012}) the density
distributions of the largest and the smallest are given by

\begin{equation}
F_U(n,x)=\frac{\rho(x)}{2}\left[1+(-1)^n\left( \mbox{erf}(x)-
\frac{x\exp(-x^2)}{\sqrt{\pi}}\right)\right] .
\end{equation}
In Fig. 3, these distributions are shown together with the 
density. One can clearly see the reduction produced by the correlations
in the range of fluctuations of the eigenvalues. 

Turning now to the case of large matrices, by using the 
(\ref{202q}), (\ref{101v}) and (\ref{101b}) we rewrite  (\ref{101}) 
as

\begin{equation}
\rho(x)=\sqrt{\frac{N}{2}}\left[\phi_N^\prime (x)\phi_{N-1} (x)-
\phi_N (x)\phi_{N-1}^\prime (x)\right] . \label{105}
\end{equation} 
If the harmonic oscillator functions are expressed in terms of
amplitude and phase as

\begin{equation}
\phi_n(x)=A_n (x)\cos\theta_n (x)
\end{equation}
then, when substituted in (\ref{105}), the density becomes
a sum of cosines and sines whose arguments are either addition or subtraction
of the function phases. We observe that as the function $\phi_n (x)$ has $n$ zeros 
its phase $\theta_n (x)$ must have a range of variation of order $n\pi.$ Therefore, 
by subtracting two adjacent phases we get a function whose variation
is smaller than $\pi$ producing therefore no wiggles.
We can then argue that the smooth part of the 
density, $\rho_s (x) $, is obtained by collecting the terms  
in which the phases subtract while, the 
fluctuating part, $\rho_f (x)$, comes from those in which they
add. Explicitly, this leads to the expressions

\begin{equation}
\rho_s =\sqrt{\frac{N}{8}}A_N A_{N-1}\left[\left(\frac{A_N^\prime}
{A_{N}} -\frac{A_{N-1}^\prime}{A_{N-1}}\right)
\cos(\theta_N -\theta_{N-1})- (\theta^\prime_N +
\theta^\prime_{N-1})
\sin(\theta_N -\theta_{N-1}) \right] \label{2222a}
\end{equation} 
and 

\begin{equation}
\rho_f =\sqrt{\frac{N}{8}}A_N A_{N-1} \left[
\left(\frac{A_N^\prime}{A_{N}} -\frac{A_{N-1}^\prime}{A_{N-1}}\right)
\cos(\theta_N +\theta_{N-1})- (\theta^\prime_N -
\theta^\prime_{N-1})
\sin(\theta_N +\theta_{N-1})  \right] \label{2222b}
\end{equation} 
(all quantities  $\theta_n$ 's and $A_n$ 's above and others below 
are $x$-dependent, but to keep the notation less heavy, this dependence is 
often dropped). An asymptotic expression for these two density terms can be deduced
from the semi-classical formalism described in appendix A. 
We find that, asymptotically, the harmonic oscillator function is given
by 

\begin{equation}
\phi_n (x)=\sqrt{\frac{2}{\pi}}\frac{\cos\left[\xi_n(x)-n/2\right] \pi}
{\left(2n+1-x^2\right)^{1/4}}, \label{2222c}
\end{equation}
where

\begin{equation}
\xi_n (x)=\frac{2n+1}{2\pi}\left[\arcsin(\frac{x}{\sqrt{2n+1}})
+\frac{x}{\sqrt{2n+1}}\sqrt{1-\frac{x^2}{2n+1}}\right] \label{116b}
\end{equation} 
is the classical mechanical action obtained integrating 
the  momentum $p=\sqrt{2n+1-x^2}.$ In (\ref{2222c}), the phase
$n\pi/2$ has been introduced to fix the
parity of the function. 
Eq. (\ref{2222c}) together with
(\ref{116b}) determine the phases and amplitudes in  (\ref{2222a}) and (\ref{2222b}) 
Substituting them and neglecting the
derivatives of the amplitude, we find, after neglecting higher order
terms, that the density takes the simple analytic form

\begin{equation}
\rho(x)=\rho_W (x)-\frac{\sqrt{2N}\cos[N-2\xi(x)]\pi}{2\pi^3\rho^2_W(x)},
 \label{115}
\end{equation}
where 

\begin{equation}
\rho_W (x)=\left\{ 
\begin{array}{rl}
\frac{1}{\pi} \sqrt{2N-x^2} , & 
\mid x\mid<\sqrt{2N} \\ 
0, & \mid x \mid > \sqrt{2N}  \\ 
\end{array}
\right. \label{115b}
\end{equation}
is  Wigner's semi-circle law and 

\begin{equation}
\xi (x)=\left\{
\begin{array}{rl}
-\frac{N}{2}, & x  < -\sqrt{2N}  \\ 
\frac{N}{\pi}\left[\arcsin(\frac{x}{\sqrt{2N}})
+\frac{x}{\sqrt{2N}}\sqrt{1-\frac{x^2}{2N}}\right] , &
\mid x\mid<\sqrt{2N} \\ 
\frac{N}{2}, & x  > \sqrt{2N}  \\ 
\end{array}
\right. .
\end{equation}
In deriving these equations, $N$ was assumed to be large enough  
to treat indexed quantities as continuous functions of the 
indices in such a way that the approximations
$f(N) +f(N-1)=2f(N-1/2)$ and $f(N) -f(N-1)=f^{\prime}(N-1/2)$ can
be made.   

Eq. (\ref{115}) shows that averaging out the wiggles the fluctuating 
term vanishes and the density becomes the semi-circle. 
The quantity $\xi(x)$ is the average stair-case function from which 
the so-called unfolded spectrum is calculated. This transformation
leads to a new spectrum with density  

\begin{equation}
\rho(\xi)=\frac{\rho(x)}{\rho_W (x)}
=1-\frac{\sqrt{2N}\cos(N-2\xi)\pi}{2\pi^3\rho^3_W(x)} \label{175}
\end{equation}
whose average is equal to one. Since $\xi (x)$ is a counting function,
in this variable the wiggles are equally spaced with an unit distance 
between them.
In Fig. 4, the density calculated using the approximated expression 
Eq. (\ref{115}) is compared with the density calculated with
Eq. (\ref{101}). One can see that with the exception of the wiggles 
at the very edges, those at the
bulk are well described by the asymptotic density. 

Eq.  (\ref{175}) gives analytical expressions for the smooth
and the fluctuating parts of the density which Fig. 4 shows are very 
precise at the bulk of the spectrum. To extend this precision up to 
the edge, removing the singularities at the classical turning points,
an improvement is needed to go beyond the 
asymptotic expression Eq. (\ref{2222c}). To this end, one must also 
consider the exact second
independent solution of the equation expressed as

\begin{equation}
\tilde{\phi}_n(x)=A_n (x)\sin\theta_n (x).
\end{equation}
From the pair of independent solutions, modulus and phase of the 
functions can be extracted. This second independent solution  
can be determined by integrating from the origin, $x=0,$
the differential equation with initial conditions provided 
by  the Wronskian relation

\begin{equation}
W({\phi}_n,\tilde{\phi}_n)=\phi_n (x)\tilde{\phi}^{\prime}_n (x) -
\phi^{\prime}_n (x) \tilde{\phi}_n (x) = 2/\pi. \label{1175} 
\end{equation}
In fact, since solutions of the equation can be constructed with 
a definite parity, for even $n,$ we take $\tilde{\phi}_n (x)$ to be odd 
such that $\tilde{\phi}_n (0) =0 $ and, from (\ref{1175}), 
$\tilde{\phi}^{\prime}_n (0) =2/\pi\phi_n (0).$ 
Inversely, for $n$ odd, $\tilde{\phi}_n (x)$ is taken to be even, 
$\tilde{\phi}^{\prime}_n (0) = 0$ and, from  (\ref{1175}),
$\tilde{\phi}_n (0) =- 2/\pi\phi^{\prime}_n (0) $  .

Once the pair of solutions is determined, modulus and phase are 
obtained as

\begin{equation} 
A_n  = \sqrt{\phi_n^2+\tilde{\phi}_n^2} 
\end{equation}
 and 

\begin{equation}
\theta_n  = \arctan (\tilde{\phi}_n/\phi_n) 
\end{equation}
 and their derivatives 
as

\begin{equation}
A_n A_n^{\prime}=\phi_n\phi_n^{\prime} 
+\tilde{\phi}_n\tilde{\phi}_n^{\prime} 
\end{equation}
and

\begin{equation}
\theta_n^{\prime}= \frac{2}{\pi} A_n^{-2}.
\end{equation}
The exact expression of the scaled density becomes

\begin{equation}
\rho=\frac{\rho(x)}{\rho_s (x)}=1+\frac{B\cos(\theta_N 
+\theta_{N-1} +\theta)}{\rho_s}, \label{175c}
\end{equation}
where

\begin{equation}
B = \sqrt{\frac{N}{8}}A_N A_{N-1}\sqrt{
\left(\frac{A_N^\prime}{A_{N}} -\frac{A_{N-1}^\prime}
{A_{N-1}}\right)^2 + (\theta^\prime_N -
\theta^\prime_{N-1})^2 }
\end{equation}
and

\begin{equation}
\theta = \arctan\left[\frac{A_NA_{N-1}(\theta^\prime_N -\theta^\prime_{N-1})}
{A_{N-1}A_N^\prime -A_{N} A_{N-1}^\prime}\right].
\end{equation}

Guided by intuition and numerics, let us make the ansatz that the 
eigenvalue density in the scaled variable can be decomposed as

\begin{equation}
\rho (\nu)=\sum_{k=1}^{N}\frac{1}{\sqrt{2\pi\sigma^2}}
\exp\left[-\frac{(\nu-\nu_k)^2}{2\sigma^2}\right] , \label{111}
\end{equation}
where,  according to Eq. (\ref{115}),

\begin{equation}
\nu_k=\frac{N+1}{2}-k \label{125}.
\end{equation}
with $k=1,2,...,N.$ Since the quantities $\nu$ and $\sigma$ 
may depend on position, each term in the above sum is not a true
Gaussian, however  they can be considered as nearly Gaussian 
distributions as that dependence is expected to be weak. In order to 
determine this dependence, we turn the summation in  (\ref{111}) 
into an infinite sum and rewrite it as

\begin{equation}
\rho (\nu)=\sum_{-\infty}^{\infty}\frac{1}{\sqrt{2\pi\sigma^2}}
\exp\left[-\frac{(\nu-\nu_k)^2}{2\sigma^2}\right]-R, \label{135}
\end{equation}
where

\begin{equation}
R=\sum_{k=-\infty}^{0}\frac{1}{\sqrt{2\pi\sigma^2}}
\exp\left[-\frac{(\nu-\nu_k)^2}{2\sigma^2}\right]+
\sum_{k=N+1}^{\infty}\frac{1}{\sqrt{2\pi\sigma^2}}
\exp\left[-\frac{(\nu-\nu_k)^2}{2\sigma^2}\right].
\end{equation}
This $R$ quantity is expected to affect only the tail of the 
distribution of the extreme eigenvalues which, as explained in the
beginning of this section, practically coincide with the total density. 
Therefore, this reminder can be neglected even for relatively small 
values of $N.$ 
The first term in the right hand side of (\ref{135}) can be
transformed into an infinite sum of integrals using the Poisson sum formula

\begin{equation}
\sum_{n=-\infty}^{\infty}f(t+n)=\sum_{m=-\infty}^{\infty}F(2\pi m)
\exp(2\pi \mbox{i} mt),
\end{equation}
where $F(s)$ is the Fourier transform 

\begin{equation}
F(s)=\int_{-\infty}^{\infty}f(t)
\exp(-\mbox{i} st) .
\end{equation}
Then, after performing all integrals we obtain  

\begin{equation}
\rho(\nu)=1+2\sum_{m=1}^{\infty}(-1)^m
\exp\left[-2(\pi m\sigma)^2\right]\cos\left[2\pi m(\nu+\frac{N}{2})\right],
\label{151}
\end{equation}
where the term  $R$ in (\ref{135}) was neglected. Due the presence 
of the exponential factor, the sum is dominated by its first term  
$m=1.$ Comparing this term with the oscillating term in  
Eq. (\ref{175}), we find that the phase $\nu$ and the variance
$\sigma$ depend on their positions as

\begin{equation}
\nu(x)= \frac{ \theta_N + \theta_{N-1} + \theta -\pi}{2\pi} - \frac{N}{2}
\end{equation}
and

\begin{equation}
\sigma^2(x)=-\frac{1}{2\pi^2}\ln\left[\frac{B(x)}{2\rho_s (x)}\right].
\label{155a}
\end{equation}
At the bulk of the spectrum, the phases take their 
asymptotic values $\theta_n=(\xi_n-n/2)\pi$ and $\theta =\pi/2$ such
that $\nu(x)=\xi(x)$ and  

\begin{equation}
\sigma^2(x)=\frac{3}{2\pi^2}\ln\left[\frac{\pi\sqrt{2}
\rho_W(x)}{N^{1/6}}\right].
\label{155}
\end{equation}
Once the variances have been determined, the decomposed density in the
actual spectrum variable is 

\begin{equation}
\rho(x)=\rho_s(x) \sum_{k=1}^{N}\frac{1}{\sqrt{2\pi\sigma^2(x)}}
\exp\left[-\frac{\left(\nu(x)-\nu_k\right)^2}{2\sigma^2(x)}\right] , \label{111b}
\end{equation}

In Fig. 5, these $N$ individual distributions are compared with the 
distributions obtained by performing numerical simulations for
matrices of size $N=20$ with a very good agreement. In Fig. 6, 
these density distributions are compared with those of uncorrelated
variables with the same density exhibiting the great 
effect of the correlations. 

Motivated by the good agreement between simulations and nearly
Gaussian distributions, we compare the  nearly
Gaussian distribution at the edge with Tracy and
Widom's prediction for the largest eigenvalue.
They proved that, when $N\rightarrow \infty,$ 
in a new variable $s$ defined by the linear relation 

\begin{equation}
x=\sqrt{2N}+\frac{s}{2^{1/2} N^{1/6}} , \label{817}
\end{equation}
the distribution probability, $E_2 (s)$, of the largest eigenvalue 
of the unitary ensemble, $\beta = 2,$ is given by \cite{TW}

\begin{equation}
E_{2}(s) =\exp\left[-\int_{s}^{\infty}(x-s)q^{2}(x)dx\right]
\label{148}
\end{equation}
where $q(s)$ satisfies the  Painlev\'{e} II equation

\begin{equation}
q^{\prime\prime} = s q +2q^3
\label{149}
\end{equation}
with boundary condition 

\begin{equation}
q(s)\sim \mbox{Ai}(s) \mbox{  when  } s 
\rightarrow \infty, \label{418}
\end{equation} 
where $ \mbox{Ai}(s)$ is the Airy function. 

In Fig. 7, the
distribution of the largest eigenvalue of matrices of 
size $N=20$ obtained performing numerical simulation is compared 
with both: Tracy-Widom density distribution, $E^{\prime}_2 (s)$ and 
our nearly Gaussian distribution. Both give a reasonable fit although
$N=20$ can be considered a relatively small size. In Table 1, 
the cumulants of the two distributions are shown\cite{TW1} and, as one
would expect, the figures point that ours is indeed more normal.  

\vskip 1cm
\begin{center}
\begin{tabular}{|c|c|c|c|c|}
\hline
&\mbox{mean}&\mbox{variance}&\mbox{skewness}&\mbox{kurtosis}\\ \hline
Tracy-Widom &  -1.77109  & 0.9018 & 0.224 & 0.093 \\ \hline
nearly Gaussian & -1.829 & 0.9066  & 0.114 & 0.074 \\\hline 
\end{tabular}
\vskip 0.50cm
\text{Table 1: Cumulants for the unitary case}
\end{center}

\subsubsection{Orthogonal ensemble ($\beta=1$)}

For the orthogonal case, we have to add to (\ref{1002}) the term

\begin{equation}
\gamma (x)= 
\sqrt{\frac{N}{2}} \phi_{N-1}(x)\int_{-\infty}^{\infty}dt\frac{1}{2}
 \mbox{sgn }(x-t)\phi_{N}(t)= \sqrt{\frac{N}{2}} \phi_{N-1}(x)
 \int_{0}^{x}dt\phi_{N}(t)                   ,   \label{10}
\end{equation}
where $\mbox{sgn }(x)$ is the sign function, 
if $N$ is even, and a further term

\begin{equation}
\phi_{N}(x)/\int_{-\infty}^{\infty}\phi_{N}(t)dt ,
\end{equation}
if $N$ is odd\cite{Mehta}.

As done in the unitary case, it is instructive to start discussing the decomposition
of the spectral density of matrices of size $N=2.$ This will
illustrate
 the differences 
between the unitary and the  orthogonal cases.  Evaluating the integral in (\ref{10}) 
we find that the extra term is

\begin{equation}
\gamma (x)=-\frac{2x^2\exp(-x^2)}{\sqrt{\pi}}+\frac{\exp(-x^2/2)x}{\sqrt{\pi}}
\int_{0}^{x}dt\exp(-\frac{t^2}{2})  .
\end{equation}
Adding this term to (\ref{1012}), its first term cancels the second
term in (\ref{1012}) and the density becomes

\begin{equation}
\rho(x)=\frac{\exp(-x^2)}{\sqrt{\pi}} + \frac{\exp(-x^2/2)}{\sqrt{2}}
\mbox{erf} (\frac{x}{\sqrt{2}})  \label{1012b} .
\end{equation}
The important point is that this cancellation of terms removes the
wiggles in the unitary 
density in such a way that the orthogonal becomes a flat function. 
The individual distributions 
of the two eigenvalues are easily calculated to be given by 

\begin{equation}
F(0,x)=\frac{\sqrt{2}\exp(-x^2)}{4}\mbox{erfc}(-\frac{x}{\sqrt{2}}) + 
\frac{\exp(-x^2)}{2\sqrt{\pi}}
\end{equation}
for the greater while  distribution of the smaller is obtained by
subtracting the above $F(0,x)$ from Eq. (\ref{1012b}). 
For the uncorrelated case, with $n=0,1$ and $\rho(x)$ given by 
(\ref{1012b}), the density distributions are

\begin{equation}
F_U(n,x)=\frac{\rho(x)}{2}\left[1+(-1)^n\left( \mbox{erf}(x)-
\frac{\exp(-x^2/2)}{\sqrt{2}} \mbox{erf}(\frac{x}{\sqrt{2}})\right)\right] .
\end{equation} 
Fig. 8 shows the spectral density of matrices of size $N=2$ of the orthogonal 
ensemble decomposed in eigenvalue individual contributions. In
contrast to the unitary case , Fig. 3, in the orthogonal case, the two
individual contributions add in such a way that density becomes flat 
at the top.     

Turning now to the case of matrices of large sizes, we have to calculate the integral 

\begin{equation}
I_N (x)=  \int_{0}^{x}dt\phi_{N}(t)  .
\end{equation}
Starting with the derivation of an asymptotic expression for it,  we use the
differential equation satisfied by the $\phi_N (x)$ functions to
rewrite the it as 

\begin{equation}
I_N (x)= - \int_{0}^{x}dt\frac{\phi^{\prime\prime}_{N}(t)}{2N+1-t^2}.  
\end{equation}
We recall that the denominator in the above integrand is the square of
the classical momentum supposed 
to be large. Therefore integration by parts can be used to obtain a
series in inverse powers of the momentum or, equivalently, of the
density, whose the first two terms are

\begin{equation}
I_N (x)= - \frac{\phi^{\prime}_{N}(x)}{2N+1-x^2}+ \frac{2x\phi_{N}(x)}{(2N+1-x^2)^2} .  
\end{equation}
Substituting this term in Eq. (\ref{10}) and replacing the functions
by their semi-classical 
approximation, we find that the first additional extra term cancels
the oscillating term of the 
unitary case. This canceling makes necessary to take into account 
higher order terms, which can be done using the expansion 

\begin{equation}
(2N \pm 1 - x^2)^\mu = (2N-x^2)^\mu  \pm \mu(2N-x^2)^{\mu-1} .
\end{equation}
By doing this, we end up with the following asymptotic expression for 
the density

\begin{equation}
\rho (x)= \rho_W  - \frac{1}{2\pi^2\rho_W} + 
\frac{\sqrt{2N}}{2\pi^5\rho^{4}_W}\cos\left(2\xi -\frac{N}{2}\right)\pi  
+ \frac{3x\sqrt{2N}}{8\pi^6\rho^{5}_W}\sin\left(2\xi -\frac{N}{2}\right)\pi \label{15g} 
\end{equation}
in the orthogonal case. The first two terms in (\ref{15g}) correspond to the smooth 
part of the density while the two last ones to its fluctuating part. 

In order to go beyond this asymptotic expression, removing, as done in
the unitary case, its singularities, we assume, that the integral $I_N
(x)$ can be written as

\begin{equation}
I_N (x) =Q_1 (x)\cos\theta_N (x)+ Q_2 (x) \sin\theta_N (x)  , \label{44}
\end{equation}
where  $\theta_N (x) $ is the phase of the function $\phi_N (x)$ and
$Q_1 (x)$ and  $Q_2 (x)$ are smooth functions of the position.
These functions can be determined considering the function
${\tilde I}_N (x) $ related to the second independent solution  
as ${\tilde I}^{\prime}_N (x) ={\tilde \phi}_{N}(x). $  
From the asymptotic analysis, we deduce that it satisfies, at the
origin, the condition 

\begin{equation}
{\tilde I}_N (0) =-\frac{ {\tilde \phi}_{N}^{\prime}(0)}{2N+1} .
\end{equation} 
and can be written as  

\begin{equation}
{\tilde I}_N (x) =Q_1 (x)\sin\theta_N (x)- Q_2 (x) \cos\theta_N (x).
\label{44f}  
\end{equation} 
Eqs. (\ref{44}) and (\ref{44f})  can be inverted to give

\begin{equation}
Q_1 (x) =I_N (x)\cos\theta_N (x) + {\tilde I}_N(x)\sin\theta_N (x) ,
\end{equation}
and

\begin{equation}
Q_2 (x) =I_N (x)\sin\theta_N (x) - {\tilde I}_N(x)\cos\theta_N (x)  .
\end{equation}
Once these two functions are determined the integral Eq. (\ref{44}) is
expressed in terms of amplitudes and phase and can be replaced in the 
additional term $\gamma(x)$ which can also be decomposed as a sum of
smooth and fluctuating terms as  

\begin{equation}
\gamma_s =\sqrt{\frac{N}{8}}A_{N-1}\left[Q_1
\cos(\theta_N -\theta_{N-1})+ Q_2
\sin(\theta_N -\theta_{N-1}) \right] \label{1111a}
\end{equation} 
and 

\begin{equation}
\gamma_f =\sqrt{\frac{N}{8}} A_{N-1} \left[Q_1
\cos(\theta_N +\theta_{N-1})+ Q_2
\sin(\theta_N +\theta_{N-1})  \right]. \label{1111b}
\end{equation} 

By adding $\gamma_s$ and $\gamma_f$ to Eqs. (\ref{2222a})
and (\ref{2222b}) respectively the scaled density is

\begin{equation}
\rho=\frac{\rho(x)}{\rho1_s (x)}=1+\frac{B\cos(\theta_N 
+\theta_{N-1} +\theta)}{\rho_{1s}}, 
\end{equation}
where

\begin{equation}
\rho_{1s} (x)= \rho_s (x) + \gamma_s (x) ,
\end{equation}

\begin{equation}
B = \sqrt{\frac{N}{8}}A_N A_{N-1}\sqrt{
\left(\frac{A_N^\prime}{A_{N}} -\frac{A_{N-1}^\prime}
{A_{N-1}}+\frac{Q_1}{A_N}\right)^2 + \left(\theta^\prime_N -
\theta^\prime_{N-1}-\frac{Q_2}{A_N}\right)^2 }
\end{equation}
and

\begin{equation}
\theta = \arctan\left[\frac{\theta^\prime_N -\theta^\prime_{N-1}
-Q_2/ A_N}
{A_N^{\prime}/A_N - A_{N-1}^{\prime}/A_{N-1}+Q_1/A_N} \right].
\end{equation}

The formalism developed in the unitary case still applies and, with
the above expressions, the scaled variable and the variance are
deduced to be given by 

\begin{equation}
\nu(x)=\frac{\theta_{N}+\theta_{N-1}+\theta-\pi}{2\pi}-\frac{N}{2}
\end{equation}
 and
 
\begin{equation}
\sigma^{2}(x)=-\frac{1}{2\pi^{2}}\ln\left[\frac{B(x)}{2\rho_{1s}(x)}\right].
\label{15h}
\end{equation}
At the bulk, from Eq. (\ref{15g}), the scaled variable becomes

\begin{equation}
\nu(x)=\xi(x)+\arctan\left[\frac{3x}{4\pi\rho_{W}(x)}\right]
\end{equation}
and the density smooth term and the amplitude are given by
 
\begin{equation}
\rho_{1s}(x)=\rho_{W}(x)-\frac{1}{2\pi^{2}\rho_{W}(x)}
\end{equation}
and

\begin{equation}
B(x)=\frac{\sqrt{2N}}{2\pi^{5}\rho_{W}^{4}}\sqrt{1+\frac{9x^{2}}{16\pi^{2}
\rho_{W}^{2}}}
\end{equation}
respectively, which replaced in Eq. (\ref{15h}) gives the asymptotic
expression of the variance.

In Fig. 9, the  density distributions of individual eigenvalues of matrices 
of size $N=20$ are shown together with distributions obtained
performing numerical simulations. The result is good specially at the
bulk of the spectrum.  
 
For the largest eigenvalue, Tracy and Widom predict that for
the orthogonal ensemble ($\beta=1$), the probability distribution 
function in the same scaled variable $s$ of the unitary case
is given by

\begin{equation}
\left[E_{1} (s)\right]^2 =E_{2}(s)\exp\left[-\mu (s)\right]
\end{equation}
where 

\begin{equation}
\mu (s) =\int_{s}^{\infty} q(x)dx.
\end{equation}
In Fig. 10, this prediction is compared with our nearly Gaussian
density distribution and with results of simulations. It is clear that
the nearly Gaussian distribution give a better description. However, 
comparing the cumulants\cite{TW1} depicted in Table 2, we see that
actually, apart from a
shift to the right of the Tracy-Widom distribution, the two
distributions are quite alike.  

\vskip 1cm
\begin{center}
\begin{tabular}{|c|c|c|c|c|}
\hline
&\mbox{mean}&\mbox{variance}&\mbox{skewness}&\mbox{kurtosis}\\ \hline
Tracy-Widom &  -1.20653  & 1.2580 & 0.293 & 0.165 \\ \hline
nearly Gaussian & -1.382  & 1.264  & 0.325 & 0.067 \\\hline 
\end{tabular}
\vskip 0.50cm
\text{Table 2: Cumulants for the orthogonal case}
\end{center}

\subsection{Wishart matrices}

Consider a rectangular matrix $X$ of size ($M$x$N$) whose elements 
are sorted independently from a Gaussian distribution, a Wishart
square matrix $W$ of size $N$ is then defined by taking the product  
$W=X^{\dagger}X$. It can be shown that for $M\geq N$, the joint 
probability distribution of the positive eigenvalues of the random 
matrices $W$, for the three symmetry classes, are given by 

\begin{equation}
P(x_1,x_2,...,x_N)=K_N \exp(-\frac{\beta}{2}\sum_{k=1}^N x_k)
\prod_{i=1}^{N} x_i^{\frac{\beta}{2}(1+M-N)-1}
\prod_{j>i}\mid x_j-x_i\mid^\beta .
\end{equation}
From (\ref{1001}), the positive eigenvalues of this ensemble are
confined by the potential $V(x)=x-(1+M-N-2/\beta)\log(x)$ and the
polynomials are the generalized Laguerre polynomials.

For the unitary case, $\beta = 2, $ the eigenvalue density is  

\begin{equation}
\rho(x)=\sum_0^{N-1} \psi_n ^2(x),\label{771}
\end{equation}
where, with $\alpha=M-N,$

\begin{equation}
\psi_n^{\alpha}(x)= \sqrt{\frac{n!}{(n+\alpha)!}}
\exp(-x/2)x^{\alpha/2}L_{n}^{\alpha}(x) \label{771a}
\end{equation}
in which the $L_{n}^{\alpha}(x)$ are the Laguerre polynomials

\begin{equation}
L_{n}^{\alpha}(x)= \sum_0^{n}(-1)^j \frac{(n+\alpha)!x^j }{(n-j)!
(\alpha+j)!j!}.\label{77b} 
\end{equation}
From (\ref{77b}), we derive 

\begin{equation}
\frac{dL_{n}^{\alpha}}{dx}=-L_{n-1}^{\alpha+1}
\end{equation}
which used in (\ref{202q}) yields 

\begin{equation}
\rho(x)=\exp(-x) x^{\alpha} \frac{\Gamma(N)}{\Gamma(N+\alpha)}
\left[ L_{N-1}^\alpha  (x)L_{N-1}^{\alpha+1} (x)-
 L_{N}^\alpha  (x)L_{N-2}^{\alpha+1} (x) \right].
\end{equation}
Inverting (\ref{771a}) to express the polynomials in terms of the 
$\psi$-functions, the density becomes 

\begin{equation}
\rho(x)=\sqrt{\frac{N(N+\alpha)}{x}}
\left[ \psi_{N-1}^\alpha  (x) \psi_{N-1}^{\alpha+1} (x)-
\sqrt{\frac{N-1}{N}}
\psi_{N}^\alpha  (x)\psi_{N-2}^{\alpha+1} (x) \right].\label{771c}
\end{equation}
Finally the derivative of Eq. (\ref{771a}) gives the relation

\begin{equation}
\frac{d\psi_{n}^{\alpha}}{dx}=\left(\frac{\alpha-x}{2x}\right)
\psi_{n}^{\alpha}-\sqrt{\frac{n}{x}}\psi_{n-1}^{\alpha+1}
\end{equation}
which used in  (\ref{771c}) allows to write the density as

\begin{equation}
\rho(x)=\sqrt{N(N+\alpha)}
\left(\psi^{\alpha}_{N}\frac{d}{dx}\psi _{N-1}^{\alpha}-
\psi_{N-1}^\alpha\frac{d}{dx}\psi_{N}^{\alpha}\right) 
\end{equation}
an expression now ready for our analysis. We start by remarking that 
the $\psi_n(x),$ in appropriate units, 
are the functions which satisfy the Schr\"{o}dinger
wave equation of the hydrogen atom 

\begin{equation}
x\frac{d^2 \psi}{dx^2}+\frac{d \psi}{dx}+\left[n+
\frac{1}{2}-\frac{(\alpha-x)^2}{4x}\right]\psi=0 \label{444}
\end{equation}
(from now on, the superscript $\alpha$ will be omitted).
As in the Wigner case, we write these functions in terms of amplitude
and phase as 

\begin{equation}
\psi_n (x)= A_n (x)\cos\theta_n (x)
\end{equation}
and substitute, in the expression of the 
density, the two functions to extract its smooth part  

\begin{equation}
\rho_s =\sqrt{\frac{N(N+\alpha)}{4}}
A_N A_{N-1}\left[\left(\frac{A_{N-1}^\prime}
{A_{N-1}} -\frac{A_{N}^\prime}{A_{N}}\right)
\cos(\theta_N -\theta_{N-1})+ (\theta^\prime_N +
\theta^\prime_{N-1})
\sin(\theta_N -\theta_{N-1})  \right]
\end{equation} 
and its fluctuating part 

\begin{equation}
\rho_f =\sqrt{\frac{N(N+\alpha)}{4}}
A_N A_{N-1} \left[
\left(\frac{A_{N-1}^\prime}{A_{N-1}} -\frac{A_{N}^\prime}{A_{N}}\right)
\cos(\theta_N +\theta_{N-1})+ (\theta^\prime_N -
\theta^\prime_{N-1})
\sin(\theta_N +\theta_{N-1})  \right]. 
\end{equation} 

In order to apply the semi-classical formalism, we first transform 
equation (\ref{444}) in the differential equation 

\begin{equation}
\frac{d^2 (\sqrt{x}\psi)}{dx^2}+
\frac{-x^2+2(2n+\alpha+1)x-\alpha^2+1}{4x^2} (\sqrt{x}\psi)=0 .
\end{equation}
satisfied by the function $\sqrt{x}\psi(x).$ In this equation, the
associated classical moment is

\begin{equation}
p(x)= \frac{1}{2x}\sqrt{(x-x_1)(x_2-x)}, 
\end{equation}
where

\begin{equation}
x_{1,2}= 2n+\alpha+1\mp \sqrt{ (2n+\alpha+1)^2 +1-\alpha^2 }. 
\end{equation}
and the asymptotic wave function is 

\begin{equation}
\psi_n (x)=\sqrt{\frac{2}{\pi}}\frac{\cos(\xi_n-\pi/4)}
{\left[(x_2 -x)(x-x_1)\right]^{1/4}}, \label{5555c}
\end{equation}
where

\begin{equation}
\xi_n =\frac{1}{4\pi}\left[
\begin{array}{rl}
-4\sqrt{x_{1}x_{2}}\arctan \sqrt{\frac{x_{2}(x-x_{1})}{x_{1}(x_{2}-x)}}    
+ (x_{2}+x_{1})\arccos\left(\frac{x_{2}+x_{1}-2x}
{x_{2}-x_{1}}\right)                       \\     
+ 2\sqrt{(x_{2}-x)(x-x_{1})}
\end{array}  
\right.
\end{equation}
Substituting this approximate wave
function in the smooth and the fluctuating parts of the density and
neglecting derivatives of the amplitudes, we arrive at  
the asymptotic expression

\begin{equation}
\rho(x)=\rho_{MP} (x)-\frac {x_{+}-x_{-}}  {16\pi^3 
x^{2} \rho_{MP}^2 (x) }
\cos[2\xi(x)]\pi , \label{615}
\end{equation}
where the first term is the Marchenko-Pastur density\cite{Pastur}

\begin{equation}
\rho_{MP} (x)=\frac{1}{2\pi x}
\sqrt{(x_{+}-x)(x-x_{-})}, 
\end{equation}
in which, with $c=\sqrt{\frac{M}{N}}$, $ x_{\pm}=N(c\pm 1)^2.$
Similarly, the function $\xi (x)$ appearing in the cosine argument is
the counting number function

\begin{equation}
\xi (x) =\frac{1}{4\pi}\left[
\begin{array}{rl}
-4\sqrt{x_{-}x_{+}}\arctan \sqrt{\frac{x_{+}(x-x_{-})}{x_{-}(x_{+}-x)}}    
+(x_{+}+x_{-})\arccos\left(\frac{x_{+}+x_{-}-2x}
{x_{+}-x_{-}}\right)                            \\
+ 2\sqrt{(x_{+}-x)(x-x_{-})}      
\end{array}  
\right.
\end{equation}
associated to the Marchenko-Pastur density. As in previous case,
approximations were made by treating indexed quantities as continuous
functions of their indices.
Eq. (\ref{615}) shows that the Marchenko-Pastur density plays,
for the Wishart matrices, the same r\^{o}le the Wigner's semi-circle 
does for the Gaussian ensembles and, by averaging out the wiggles 
produced by the oscillating term, it is obtained. 
We remark that the above expression for the next to leading order
term of the asymptotic, has an analog structure to that of the
Gaussian expression Eq. (\ref{175}), namely, both are, 
basically, the ratio between the superior limit and the cubic 
of the asymptotic density. As before, we use the counting function
$\xi$ as independent variable with density 

\begin{equation}
\rho(\xi)=\frac{\rho(x)}{\rho_{MP} (x)}
=1-\frac {x_{+}}  {16\pi^3 
x^{2} \rho_{MP}^3 (x) }
\cos[2\xi(x)]\pi
\end{equation}
In Fig. 11, we compare the above density approximation with its
exact expression. It is seen that it is indeed a very good
approximation with the exception of the regions near the two
borders.

As in the Wigner case, a complete decomposition of the density 
which includes eigenvalues at the border can be achieved using 
the second independent solution of the wave equation which 
can be obtained from the integral representation

\begin{equation}
\psi_{n} (x)  =\frac{ 2^{\alpha}(-1)^n }{\pi x^{\alpha/2}}
\sqrt{\frac{(n+\alpha)!}{n!}}\int_0^{\frac{\pi}{2}}
d\theta \cos^{\alpha -1}\theta \cos\left[ \frac{x\tan\theta}{2} 
-(2n + \alpha +1)\theta\right] .  \label{571}
\end{equation}
of the first solution. That this function, with $\alpha>1,$ is 
equivalent to (\ref{771a}) can be seen by changing the integration 
variable to $u=\tan\theta$ that transforms the integral in (\ref{571})
in the complex integral

\begin{equation}
\psi_{n} (x)  =
\frac{2^{\alpha}(-1)^n  }{2\pi x^{\alpha/2} }\sqrt{\frac{(n+\alpha)!}
{n!}}(-i)^{2n+\alpha+1}\int_{-\infty}^{\infty}
\frac{du\exp(\frac{iux}{2})(u+i)^n}{(u-i )^{n+\alpha+1}} 
\end{equation}
which performed by residues reproduces (\ref{771a}). 
This integral representation suggests that the other independent solution 
is provided by replacing in the integrand the cosine of the
oscillating factor by the sine. However, this is not enough, and yet 
another term has to be subtracted to construct the solution 

\begin{equation}
{\tilde \psi}_{n} (x)  =\frac{2^{\alpha}(-1)^n }{\pi x^{\alpha/2} }
\sqrt{\frac{(n+\alpha)!}{n!}} \left\{
\begin{array}{rl}\int_0^{\frac{\pi}{2}}
d\theta \cos^{\alpha -1}\theta \sin\left[ \frac{x\tan\theta}{2} 
-(2n + \alpha +1)\theta\right]   +                        \\
   \\
\int_0^{\infty}
d\theta \cosh^{\alpha -1}\theta \exp\left[ \frac{x\tanh\theta}{2} 
-(2n + \alpha +1)\theta\right]   
\end{array}  
\right.             .
\end{equation}
Now, we have a pair of independent solutions and modulus and phase are 
determined through the relations

\begin{equation}
A_n  = \sqrt{\psi_n^2+\tilde{\psi}_n^2} 
\end{equation}
and 

\begin{equation}
\theta_n  = \arctan (\tilde{\psi}_n/\psi_n)  
\end{equation}
with derivatives given by

\begin{equation}
A_n A_n^{\prime}=\psi_n\psi_n^{\prime} 
+\tilde{\psi}_n\tilde{\psi}_n^{\prime} 
\end{equation}
and, using the Wronskian $W(\psi\tilde{\psi}) (x)= \frac{2}{\sqrt{\pi}},$

\begin{equation}
\theta_n^{\prime}= \frac{1}{\pi} A_n^{-2}.
\end{equation}
In calculating these quantities the two recurrence relations 

\begin{equation}
Z_{n-2}=\frac{1}{\sqrt{(n-1)(n+\alpha-1)}}\left[
(2n+\alpha-1-x)Z_{n-1}
-\sqrt{n(n+\alpha)}Z_{n} \right]
\end{equation}
and

\begin{equation}
x\frac{dZ_{n}}{dx}=\left(\frac{2n+\alpha-x}{2}\right)Z_{n}^{\alpha}-
\sqrt{n(n+\alpha)}Z_{n-1},
\end{equation}
where $Z_n$ denotes any one of the two solutions are useful.

The precise expression for the decomposition of the density in smooth
and fluctuating parts is

\begin{equation}
\rho=\frac{\rho(x)}{\rho_s (x)}=1-\frac{B\cos(\theta_N 
+\theta_{N-1} - \theta)}{\rho_s(x)}, \label{375c}
\end{equation}
where

\begin{equation}
B = \sqrt{\frac{N(N+\alpha)}{4}}A_N A_{N-1}\sqrt{
\left(\frac{A_N^\prime}{A_{N}} -\frac{A_{N-1}^\prime}
{A_{N-1}}\right)^2 + (\theta^\prime_N -
\theta^\prime_{N-1})^2 }
\end{equation}
and

\begin{equation}
\theta = \arctan\left[\frac{A_NA_{N-1}(\theta^\prime_N -\theta^\prime_{N-1})}
{A_{N-1}A_N^\prime -A_{N} A_{N-1}^\prime}\right].
\end{equation}
 
Turning now to the decomposition of Wishart eigenvalue
density in a sum of individual nearly Gaussian distributions, we write

\begin{equation}
\rho(\nu)=\sum_0^{N-1}\frac{1}{\sqrt{2\pi\sigma^2}}
\exp\left[-\frac{(\nu-\nu_k)^2}{2\sigma^2}\right] ,
\end{equation} 
where

\begin{equation}
\nu_k=\frac{1}{2}+k .
\end{equation}
with $k=0,1,2,...,N-1.$ 
As was done in the Wigner case, we turn the above sum into an infinite
sum which can be transformed, after neglecting the rest, into the 
infinite sum

\begin{equation}
\rho(\nu)=1+2\sum_{m=1}^{\infty}(-1)^m
\exp\left[-2(\pi m\sigma)^2\right]\cos\left[2\pi m(\nu)\right],
\end{equation}
using Poisson sum formula. The dependence of the
phase and the variance on their positions along the spectrum is
then determined to be given by  

\begin{equation}
\nu(x)= \frac{ \theta_N + \theta_{N-1} - \theta}{2\pi}
\end{equation}
and

\begin{equation}
\sigma^2(x)=-\frac{1}{2\pi^2}\ln\left[\frac{B(x)}{2\rho_s (x)}\right].
\end{equation}
At the bulk, these quantities approach their asymptotic expressions
$\nu(x)=\xi(x)$ and

\begin{equation}
\sigma^2(x) = \frac{3}{2\pi^2}\ln\left[\frac{2\pi 
(2x)^{2/3} \rho_{MP} (x) }{x_{+}^{1/3}}\right] .
\end{equation}

In Fig. 12, the individual eigenvalue distributions of $N=20$ Wishart
matrices are shown. It is seen that also for this ensemble our
formalism based on nearly Gaussian distributions give account of the
results obtained performing numerical simulations. 

For the largest eigenvalue the prediction is that its
distribution follows, for large $N$, the Tracy-Widom distribution 
in the scaled variable\cite{Johnstone} 

\begin{equation}
s = \frac{x-x_+}{x_+^{2./3.}(MN)^{-1/6}}.
\end{equation}
In Fig. 13, the two theoretical predictions are compared with results
from numerical simulations. Finally, in Fig. 14, the density
distributions of the smallest eigenvalue is magnified to show that
the nearly Gaussian distribution gives a good fit to the numerical
simulation.

\section{Concluding remarks}

We have performed a decomposition in individual contributions of the 
exact density of eigenvalues of matrices of the unitary and the orthogonal 
Gaussian ensembles and of the unitary matrices of the Wishart
ensemble. This decomposition works for relatively small values of 
matrix sizes and shows that eigenvalue are well described by nearly
Gaussian distributions. As the matrices sizes increase, these
distributions tend, at the bulk of the spectra, to true Gaussian such
that the exact results of Refs. \cite{Jonas,Sean} are recovered. 
These results should apply to spectra of systems whose classical
analog are chaotic and be experimentally tested. The present analysis
should also work for ensembles connected to others orthogonal
polynomials. Recently, invariant non-ergodic ensembles have been
introduced whose ensemble densities are averages over the 
Wigner's semi-circle\cite{Pato}, in the Gaussian case, and the 
Marchenko-Pastur density\cite{wishart}, in the Wishart case. In the
Gaussian case, the effect
of non-ergodicity on the Tracy-Widom distribution has been investigated
\cite{Pato1}. It seems possible to extend this study to the individual
distributions of the eigenvalues at the spectral bulk. This is
interesting in connection with the behavior of the eigenvalues of 
the embedded ensembles\cite{Flores,Asaga}.    
On the other hand, the high 
development the experimental physics achieved in the 
field of cold atoms, should perhaps reach the point in which individual 
atoms can be observed. In this case, the behavior of each atom of a
Girardeau gas can be measured and the decomposition in individual
contributions be checked.    
 
This work is supported by a project CAPES-COFECUB and by 
the Brazilian agencies CNPq and FAPESP. 

\appendix

\section{Semi-classical wave function approximation}

Consider a wave function $\Psi(x)$ that satisfies the wave equation

\begin{equation}
\frac{d^2 \Psi }{dx^2}+p^2 (x) \Psi=0 , \label{333}
\end{equation} 
where the momentum $p(x)$ is such that $p^2 (x)>0$ in the 
interval $(x_1,x_2).$ Let us search a solution of (\ref{333})
in this interval of the form 

\begin{equation}
\Psi (x)=M(x)\cos\beta(x) . \label{333a}
\end{equation}
Substituting  (\ref{333a}) in (\ref{333}) and neglecting the second
derivative of the modulus, we find that modulus and phase satisfy
the equations

\begin{equation}
\beta^{\prime} = p(x) 
\end{equation}
and

\begin{equation}
M\beta^{\prime\prime}+2M^{\prime}\beta^{\prime}=0
\end{equation}
with solutions $\beta = \int^{x}_{x_1} dx p(x) +\beta_0$ and  
$M=Cp^{-1/2}.$ Finally the constant $C$ is fixed by the normalization
condition $\int \Psi^2 (x)dx=1$ that gives

\begin{equation}
C = \left[\frac{1}{2}\int^{x_2}_{x_1} \frac{dx\left[1+\cos 2\beta(x)\right]} 
{p(x)}\right]^{-1/2} =\left[\frac{1}{2}\int^{x_2}_{x_1} \frac{dx} 
{p(x)}\right]^{-1/2} ,
\end{equation}
where, consistently with the semi-classical method, the oscillating term 
in the integrand is neglected.

As a second order differential equation, Eq. (\ref{333}) has a second
independent solution which in the same approximation we assumed to
be given by

\begin{equation}
\tilde{\Psi} (x)=M(x)\sin\beta(x) . 
\end{equation}
These two independent solutions satisfy the Wronskian relation

\begin{equation}
W(\Psi, \tilde{\Psi}) (x)= \Psi (x) \tilde{\Psi}^{\prime} (x)
-\Psi^{\prime} (x) \tilde{\Psi} (x)
=C^2
\end{equation}
in the same order of approximation.

{\bf Figure Captions}

Fig. 1 Distributions of $N=20$ individual Gaussian variables with density
$N \exp(-x^2 )/\sqrt{\pi}$ .
 
Fig. 2 For $N=100,$  histogram of the largest Gaussian variable
scaled as $t(x)= \frac{N}{2} \mbox {erf} (x)$ is plotted together 
with the predicted Gumbel density distribution. 
 
Fig. 3 Decomposition of the $N=2$ density of the unitary
ensemble (the black line) in individual contributions: 
eigenvalues (blue lines) and uncorrelated random variables (red line).
 
Fig. 4 The relative difference $(\rho_{asym}-\rho)/\rho$ 
between the exact density $\rho$ and its asymptotic is plotted versus the
unfolded variable $\xi$.
 
Fig. 5 For the unitary case, the $N$ individual distributions calculated 
with our formalism  complete (blue line) and asymptotic (red line)
compared with the result of numerical simulations 
(black line) for $N=20.$ 
 
Fig. 6  Decomposition of the $N=20$ density of the unitary
ensemble (the black line) in individual contributions: 
eigenvalues (blue lines) and uncorrelated random variables (red line).
 
Fig. 7 Theoretical distributions of the largest of $N=20$ eigenvalues 
compared with numerical simulation (blue line): nearly Gaussian
(black line) and Tracy-Widom (blue line) and asymptotic (red line)
compared with the result of numerical simulations 
(black line).  

Fig. 8  Decomposition of the $N=2$ density of the orthogonal
ensemble (the black line) in individual contributions: eigenvalues 
(blue lines) and uncorrelated random variables (red line).
 
Fig. 9 For the orthogonal ensembles, $N=20$ individual distributions 
calculated with our formalism  complete (blue line) and asymptotic 
(red line) compared with the result of numerical simulations 
(black line).

Fig. 10 For the orthogonal ensemble, largest eigenvalue density 
distribution for $N=20$: nearly Gaussian (black line), 
Tracy-Widom (red line) and simulation (blue line).

Fig. 11 The relative difference $(\rho_{asym}-\rho)/\rho$ between 
the exact density $\rho$ and its asymptotic is plotted versus the
unfolded variable $\xi$.
 
Fig. 12  Decomposition of the $N=20$ density of the unitary
ensemble (the black line) in individual contributions: 
eigenvalues (blue lines) and uncorrelated random variables (red line).

Fig. 13 For the Wishart ensemble, largest eigenvalue density 
distribution for $N=20$: nearly Gaussian (black line), 
Tracy-Widom (red line) and simulation (blue line). 
 
Fig. 14 For the Wishart ensemble, the smallest eigenvalue density
distribution: nearly Gaussian (black line) and simulation (red line).


\begin{thebibliography}{99}

\bibitem{Wishart} J. Wishart,Biometrica {\bf 20}, 32 (1928).

\bibitem{Porter} C. S. Porter, {\it Statistical Theories of Spectra}, Academic 
Press, New York (1965).

\bibitem{Mehta}  M. L. Mehta, {\it Random Matrices}
(Elsevier Academic Press, 3nd Ed., 2004).

\bibitem{Gira} M. Girardeau, J. Math. Phys. {\bf 1}, 516 (1960).

\bibitem{Kolo} E. B. Kolomeisky, T. J. Newman, J. P. Straley, 
and X. Qi, Phys. Rev. Lett {\bf 85}, 1146 (2000).

\bibitem{TW} C. A. Tracy and H. Widom, Commun. Math. Phys. {\bf 159},
 151 (1994) and  {\bf 177}, 727 (1996).

\bibitem{Johnstone} K. Johansson, Comm. Math. Phys. {\bf 209}, 437 (2000);
I. M. Johnstone, Ann. Stat. {\bf 29}, 295 (2001);
P. Forrester, Nonlinearity {\bf 19}, 2989 (2006).

\bibitem{Flores} O. Bohigas and J. Flores, Phys. Lett.  {\bf 35B}, 383 (1971).

\bibitem{Jonas} J. Gustavsson, Ann. Inst. H. Poincar\'{e}-Prob. Stat.
  {\bf 41},  151 (2005).

\bibitem{Sean} S. O'Rourke, arXiv : 0909.2677v2 [math.PR].

\bibitem{Tao} T. Tao and Van Vu, arXiv : 0906.0510 [math.PR].

\bibitem{Edelman} I. Dumitriu and A. Edelman,  J. Math. Phys. 
{\bf 43}, 5830 (2002).

\bibitem{Edelman2} I. Dumitriu and A. Edelman, 
 Ann. Inst. H. Poincar\'{e}- Prob. Stat.  {\bf 41}, 1083 (2005).

\bibitem{Coles} S. Coles, {\it An Introduction to Statistical
Modeling of Extreme Values}, Springer series in statistics (2001).

\bibitem{Weibull} W. Weibull, J. Appl. Mech.-Trans. ASME 18(3),
293 (1951).

\bibitem{Gumbel} E.J. Gumbel (1958). {\it Statistics of Extremes}
(Columbia University Press, New York, 1958).

\bibitem{TW1} C. A. Tracy and H. Widom, Proceedings of the ICM,
Beijing 2002, vol. 1, 587--596.

\bibitem{Pastur} V. A. Marchenko and L. A. Pastur, Math. USSR-Sb,
{\bf 1}, 457 (1967). 

\bibitem{Pato} 
 A. C. Bertuola, O. Bohigas, and M. P. Pato,
Phys. Rev. E {\bf 70}, 065102(R) (2004);  F. Toscano, R.O. Vallejos, 
and C. Tsallis, Phys. Rev. E {\bf 69}, 066131 (2004); 
O. Bohigas, J. X. de Carvalho, and M. P. Pato, Phys. Rev. E {\bf 77}, 
011122 (2008)

\bibitem{wishart} G. Akemann and P. Vivo,
J. Stat. Mech. P09002 (2008);  A. Y. Abul-Magd, G. Akemann and P. Vivo, 
J. Phys. A: Math. Theor. {\bf 42}, 175207 (2009); G. Akemann, 
J. Fischmann and P. Vivo, Physica A {\bf 389} 2566 (2010).

\bibitem{Pato1} O. Bohigas, J. X. de Carvalho, and M. P. Pato,
Phys. Rev. E {\bf 79}, 031117 (2009).

\bibitem{Asaga} T. Asaga, L. Benet, T. Rupp, H. A. Weidenmueller, 
Europhys.Lett. {\bf 56}, 340 (2001).


\end{thebibliography}
\end{document}